516FC29Jan2025

# Simple systems of nonlinear recursions whose evolution is rather easily ascertained

Francesco Calogero
Physics Department, University of Rome "La Sapienza", Rome, Italy
Istituto Nazionale di Fisica Nucleare, Sezione di Roma 1, Rome, Italy
Istituto Nazionale di Fisica Matematica, Gruppo Nazionale di Fisica Matematica, Italy
francesco.calogero@uniroma1.it, francesco.calogero@roma1.infn.it

**Abstract**

It is shown that the solutions of certain *systems* of *nonlinear first-order recursions* with *polynomial* right-hand sides may be rather easily *ascertained*, and display interesting evolutions in their *ticking time* variable (taking *integer* values): for instance a remarkable kind of *asymptotic periodicity*.

## 1. Introduction

This paper is one more addition to a few recent papers (see Refs. [1-5]) on *nonlinear* first-order *recursions*—and *systems* of such recursions—in which various simple examples were identified, having the remarkable property to allow a *rather detailed understanding* of the behavior of their solutions (even though these examples are—to the best of my knowledge—*not* explicitly solvable in terms of standard known functions); making them therefore suitable for useful *applications*. In the last of those papers (see Ref. [5]) the simple first-order *recursion*

$$y\left(\ell+1\right)=\left[1-y(\ell)\right]^{p}\ ,\qquad (1)$$

was investigated, where $y(\ell)$ is the *dependent* variable, $\ell=0,1,2...$ is the *independent* variable (think of it as, say, a *ticking time*) and $p$ is an *arbitrary positive integer*. In the present paper the findings of Ref. [5] are used to get information on the solutions of certain *systems* of *first-order recursions*

$$x_{n}\left(\ell+1\right)=P_{p}^{(n)}\left[x_{1}\left(\ell\right),...,x_{N}\left(\ell\right)\right]\ ,\quad n=1,2,...,N\ ,\qquad (2a)$$

*featuring* $N$ new *dependent* variables $x_n\left(\ell\right)$ and, in their right-hand sides, the following $N$ *polynomials* $P_{p}^{(n)}\left[x_{1}\left(\ell\right),...,x_{N}\left(\ell\right)\right]$ *of degree* $p$:

$$P_{p}^{(n)}\left(x_{1},...,x_{N}\right)=\sum_{q=0}^{p}Q_{q}^{(n)}\left(x_{1},...,x_{N}\right)\ ;\qquad (2b)$$

where the $N\cdot(p+1)$ polynomials $Q_{q}^{(n)}\left(x_{1},...,x_{N}\right)$ are *homogeneous* of degree $q$ in the variables $x_1,...,x_N$,

$$Q_{q}^{(n)}\left(x_{1},...,x_{N}\right)=\sum_{m_{1},m_{2},...,m_{N}=0,\ m_{1}+m_{2}+...+m_{N}=q}^{q}\left[c_{m_{1}...m_{N}}^{(n,q)}\left(x_{1}\right)^{m_{1}}...\left(x_{N}\right)^{m_{N}}\right].\qquad (2c)$$

**Notation**. Above and hereafter $N$ is a (generally *not too large*) *positive* integer, the *integer* index $n$ runs from 1 to $N$, $p$ is a *positive integer*, the *integer* index $q$ runs from 0 to $p$, the $N$ indices $m_1, m_2, ..., m_N$ each run from 0 to $q$, and the ("ticking time") *independent* variable $\ell$ runs over the *nonnegative integers* from 0 to $\infty$; and it is generally assumed that the coefficients $c^{(n,q)}_{m_1 ... m_N}$ of the polynomials $Q^{(n)}_q(x_1, ..., x_N)$ are *real* numbers, see eq. (2c); of course some of them might *vanish*: indeed the purpose of this paper is just to identify *restrictions* on these coefficients $c^{(n,q)}_{m_1 ... m_N}$, which are *sufficient* to make the systems of recursions (2) easily manageable, namely such that the behavior of its solutions is *predictable* (even if *not quite explicitly* in terms of *known* functions). ■

## 2. A change of variables

It is now convenient to introduce the following $N$ *copies* of the recursion (1),

$$y_n(\ell) = [1 - y_n(\ell)]^p \ \ , \ n = 1, 2, ..., N \ ; \tag{3}$$

the behavior of the solutions $y_n(\ell)$ of each of these *recursions* is *ascertainable* via the findings of Ref. [5]. Then clearly a standard way to identify *new* systems of *nonlinear recursions* whose evolution is as well *ascertainable,* is via the introduction of a new set of $N$ *dependent* variables $x_n(\ell)$ *linearly* related to the $N$ variables $y_n(\ell)$:

$$x_n(\ell) = a_n + \sum_{m=1}^{N} [b_{nm} y_m(\ell)] \ , \tag{4}$$

where the $N + N^2 = N(N+1)$ coefficients $a_n$ and $b_{nm}$ are assumed to be all *real* (*$\ell$-independent*) numbers (which might then be considered to be freely *assignable* parameters); with the reasonable restriction that the determinant of the $N \times N$ matrix $\mathbf{B}$ (whose elements are the numbers $b_{nm}$) be *invertible*, namely that

$$|\det(\mathbf{B})| > 0 \ . \tag{5a}$$

This allows to also introduce the *inverse* matrix

$$\widetilde{\mathbf{B}} = \mathbf{B}^{-1} \ , \tag{5b}$$

whose matrix elements $\widetilde{b}_{nm}$ are given by well-known *standard* formulas (hence no need to report them here) in terms of the elements $b_{nm}$ of the matrix $\mathbf{B}$. Then of course (4) implies

$$y_n(\ell) = \sum_{m=1}^{N} \left\{ \widetilde{b}_{nm} [x_m(\ell) - a_n] \right\} \ . \tag{6}$$

Hence from (4), (3) and (6) we get

$$x_n(\ell+1) = a_n + \sum_{m=1}^{N} \left[ b_{nm} y_m(\ell+1) \right]$$

$$= a_n + \sum_{m=1}^{N} \left\{ b_{nm} \left[ y_m(\ell) - 1 \right]^p \right\}$$

$$= a_n + \sum_{m=1}^{N} \left\{ b_{nm} \left( \sum_{m'=1}^{N} \left\{ \tilde{b}_{mm'} \left[ x_{m'}(\ell) - a_{m'} \right] \right\} - 1 \right)^p \right\} . \quad (7)$$

And it is clear that the final version of this system of recurrences is formally quite equivalent to the system of recursions (2); but it now implies an *explicit* expression of the coefficients $c^{(n)}_{m_1 ... m_N}$ in terms of the $N + N^2$ parameters $a_n$ and $b_{nm}$.

The main purpose of this paper is to make these relations more *explicit.* It follows analogous recent contributions devoted to *simple recursions*, and *systems of recursions*, whose evolution is more or less *explicitly controllable* and which may therefore be of interest in *applicative* contexts (see Refs. [2-5]). In the following, rather than considering immediately the case of an *arbitrary positive integer* value of the parameter $p$, it is wise to start from the *smaller* values of $p$. The $p = 2$ case was treated in Refs. [2], so we may begin by focusing on the next case, $p = 3$. And, whenever appropriate, we shall also focus on *smaller* values of the *integer* $N$, in order to be able to display reasonably manageable—hence therefore easily *applicable*—results.

## 3. The $\mathbf{p = 3}$ case

So in this **Section 3** we focus on the recursion

$$y(\ell+1) = \left[ 1 - y(\ell) \right]^3 \ . \quad (8a)$$

Let us recall (see [5]) that its *equilibria* are of course the roots of the *cubic* equation

$$\bar{y}^3 - 3\bar{y}^2 + 4\bar{y} - 1 = 0 \ ; \quad (8b)$$

this *cubic* equation has only 1 *real* root, namely the *irrational* number

$$\bar{y} = 1 + \left[ (\sqrt{93} - 9)/2 \right]^{1/3} \cdot 3^{-2/3} - \left[ (\sqrt{93} - 9)/2 \right]^{-1/3}$$
$$= 0.3176721961719808 \quad (8c)$$

(the other 2 roots being 2 *complex* numbers of each other, not reported here; and note that, above and hereafter, numbers written in *decimal* form are only represented *approximately*). It is moreover plain that the recursion (8a) features 2 *periodic* solutions $y(\ell)$, of period 2, one *vanishing* ($y(\ell) = 0$) for *all even* values of $\ell = 0, 2, 4, ...$, while taking the value $y(\ell) = 1$ for *all odd* values of $\ell = 1, 3, 5, ...$; the other taking instead the value $y(\ell) = 1$ for *all even* values of $\ell = 0, 2, 4, ...$, while *vanishing* ($y(\ell) = 0$) for *all odd* values of $\ell = 1, 3, 5, ...$; while *all* the other solutions starting *inside* the interval $0 < y(0) < 1$—except of course for the *equilibrium* one, $y(\ell) = \overline{y}$, see(8c)—are *asymptotically periodic* with *period* 2 as $\ell \to \infty$, jumping at *every* step from one to the other side of the equilibrium value $\overline{y}$, thereby steadily approaching the extremal values of the interval $0 < y(0) < 1$. As for the behavior—again quite easily *ascertainable*—for *initial values* $y(0)$ outside the interval $0 < y(0) < 1$, the interested reader is referred to Ref. [5].

So, the assumption is now made to have $N$ different dependent variables $y_n(\ell)$, each of them evolving according to the *cubic* recursion (8a),

$$y_n(\ell+1) = [1 - y_n(\ell)]^3 \ , \tag{9a}$$

of course each of them with a possibly *different initial* datum $y_n(0)$; while for simplicity it might—but of course it need not—be assumed that *all* these *initial* data fall *inside* the interval from 0 to 1,

$$0 < y_n(0) < 1 \ . \tag{9b}$$

Now, consider the *linear* change of variables introduced above, from the variables $y_n(\ell)$ to *new* dependent variables $x_n(\ell)$—see (7): but note that now $p = 3$—hence

$$x_n(\ell+1) = a_n + \sum_{m=1}^{N} \left[ b_{nm} y_m(\ell+1) \right] \tag{10a}$$

$$= a_n + \sum_{m=1}^{N} \left\{ b_{nm} \left[ y_m(\ell) - 1 \right]^3 \right\} \tag{10b}$$

$$= a_n + \sum_{m=1}^{N} \left\{ b_{nm} \left( \sum_{m'=1}^{N} \left\{ \tilde{b}_{mm'} \left[ x_{m'}(\ell) - a_{m'} \right] \right\} - 1 \right)^3 \right\} \tag{10c}$$

$$\begin{aligned} = \ & \sum_{m=1}^{N} \left\{ b_{nm} \left[ \left( \sum_{m'=1}^{N} \left\{ \tilde{b}_{mm'} \left[ x_{m'}(\ell) \right] \right\} \right)^3 \right. \right. \\ & -3 \left( 1 + \sum_{m'=1}^{N} \left[ \tilde{b}_{mm'} a_{m'} \right] \right) \left( \sum_{m'=1}^{N} \left[ \tilde{b}_{mm'} x_{m'}(\ell) \right] \right)^2 \\ & +3 \left( 1 + \sum_{m'=1}^{N} \left[ \tilde{b}_{mm'} a_{m'} \right] \right)^2 \sum_{m'=1}^{N} \left[ \tilde{b}_{mm'} x_{m'}(\ell) \right] \\ & \left. \left. +a_n - \left( 1 + \sum_{m'=1}^{N} \left[ \tilde{b}_{mm'} a_{m'} \right] \right)^3 \right] \right\} \ . \end{aligned} \tag{10d}$$

These are relatively explicit formulas written in terms of *elementary algebraic operations*, but they are a bit complicated—hence possibly not very useful in *applicative* contexts—as long as we consider the general case of $N$ being an *arbitrary positive integer.*

So it is wise to focus on some specific cases, beginning from $N=2$, when

$$\sum_{m'=1}^{2}\left[\widetilde{b}_{mm'}x_{m'}(\ell)\right]=\widetilde{b}_{m1}x_1(\ell)+\widetilde{b}_{m2}x_2(\ell)\ , \tag{11}$$

hence

$$x_n(\ell+1)=-a_n+b_{n1}y_1(\ell+1)+b_{n2}y_2(\ell+1) \tag{12a}$$

$$=-a_n+b_{n1}\left[y_1(\ell)-1\right]^3+b_{n2}\left[y_2(\ell)-1\right]^3 \tag{12b}$$

$$=-a_n+b_{n1}\left(\sum_{m'=1}^{2}\left[\widetilde{b}_{1m'}x_{m'}(\ell)\right]-1\right)^3+b_{n2}\left(\sum_{m'=1}^{2}\left[\widetilde{b}_{2m'}x_{m'}(\ell)\right]-1\right)^3 \tag{12c}$$

$$\begin{aligned}= \quad & -a_n+\ b_{n1}\left(\widetilde{b}_{11}x_1(\ell)+\widetilde{b}_{12}x_2(\ell)\right)^3+b_{n2}\left(\widetilde{b}_{21}x_1(\ell)+\widetilde{b}_{22}x_2(\ell)\right)^3\\ & -3b_{n1}\left(\widetilde{b}_{11}x_1(\ell)+\widetilde{b}_{12}x_2(\ell)\right)^2\ -3b_{n2}\left(\widetilde{b}_{21}x_1(\ell)+\widetilde{b}_{22}x_2(\ell)\right)^2\\ & +3b_{n1}\left(\widetilde{b}_{11}x_1(\ell)+\widetilde{b}_{12}x_2(\ell)\right)+3b_{n2}\left(\widetilde{b}_{21}x_1(\ell)+\widetilde{b}_{22}x_2(\ell)\right)-1\end{aligned} \tag{12d}$$

$$\begin{aligned}= \quad & -a_n+\ b_{n1}\left(\widetilde{b}_{11}x_1(\ell)+\widetilde{b}_{12}x_2(\ell)\right)^3+b_{n2}\left(\widetilde{b}_{21}x_1(\ell)+\widetilde{b}_{22}x_2(\ell)\right)^3\\ & -3b_{n1}\left(\widetilde{b}_{11}x_1(\ell)+\widetilde{b}_{12}x_2(\ell)\right)^2\ -3b_{n2}\left(\widetilde{b}_{21}x_1(\ell)+\widetilde{b}_{22}x_2(\ell)\right)^2\\ & +3b_{n1}\left(\widetilde{b}_{11}x_1(\ell)+\widetilde{b}_{12}x_2(\ell)\right)+3b_{n2}\left(\widetilde{b}_{21}x_1(\ell)+\widetilde{b}_{22}x_2(\ell)\right)-1\end{aligned} \tag{12e}$$

$$\begin{aligned}= \quad & -a_n+b_{n1}\,[\ \left(\widetilde{b}_{11}x_1(\ell)\right)^3+3\left(\widetilde{b}_{11}x_1(\ell)\right)^2\widetilde{b}_{12}x_2(\ell)\\ & +3\widetilde{b}_{11}x_1(\ell)\left(\widetilde{b}_{12}x_2(\ell)\right)^2+\left(\widetilde{b}_{12}x_2(\ell)\right)^3\ ]\\ & +b_{n2}\,[\ \left(\widetilde{b}_{21}x_1(\ell)\right)^3+3\left(\widetilde{b}_{21}x_1(\ell)\right)^2\widetilde{b}_{22}x_2(\ell)\\ & +3\widetilde{b}_{21}x_1(\ell)\left(\widetilde{b}_{22}x_2(\ell)\right)^2+\left(\widetilde{b}_{22}x_2(\ell)\right)^3\ ]\\ & -3b_{n1}\left[\left(\widetilde{b}_{11}x_1(\ell)\right)^2+\left(\widetilde{b}_{12}x_2(\ell)\right)^2+2\widetilde{b}_{11}x_1(\ell)\,\widetilde{b}_{11}x_1(\ell)\right]\\ & -3b_{n2}\left[\left(\widetilde{b}_{21}x_1(\ell)\right)^2+\left(\widetilde{b}_{22}x_2(\ell)\right)^2+2\widetilde{b}_{21}x_1(\ell)\,\widetilde{b}_{21}x_1(\ell)\right]\\ & +3b_{n1}\left(\widetilde{b}_{11}x_1(\ell)+\widetilde{b}_{12}x_2(\ell)\right)+3b_{n2}\left(\widetilde{b}_{21}x_1(\ell)+\widetilde{b}_{22}x_2(\ell)\right)-1\end{aligned} \tag{12f}$$

$$=Q_3^{(n)}(x_1,x_2)+Q_2^{(n)}(x_1,x_2)+Q_1^{(n)}(x,x_2)+Q_0^{(n)}(x_1,x_2)\ , \tag{12g}$$

where the following 4 *homogeneous* polynomials of degree $3, 2, 1, 0$ are introduced:

$$\begin{aligned}
Q_3^{(n)}(x_1, x_2) &= c_{30}^{(n,3)}(x_1)^3 + c_{21}^{(n,3)}(x_1)^2 x_2 + c_{12}^{(n,3)} x_1 (x_2)^2 + c_{03}^{(n,3)}(x_2)^3 \ , \\
Q_2^{(n)}(x_1, x_2) &= c_{20}^{(n,2)}(x_1)^2 + c_{11}^{(n,2)} x_1 x_2 + c_{02}^{(n,2)}(x_2)^2 \ , \\
Q_1^{(n)}(x_1, x_2) &= c_{10}^{(n,1)} x_1 + c_{01}^{(n,1)} x_2 \ , \\
Q_0^{(n)}(x_1, x_2) &= c_{00}^{(n,0)} \ ; \qquad (12h)
\end{aligned}$$

hence finally

$$\begin{aligned}
c_{30}^{(n,3)} &= b_{n1}\left(\tilde{b}_{11}\right)^3 + b_{n2}\left(\tilde{b}_{21}\right)^3 \ , \\
c_{21}^{(n,3)} &= 3\left[b_{n1}\left(\tilde{b}_{11}\right)^2 \tilde{b}_{12} + b_{n2}\left(\tilde{b}_{21}\right)^2 \tilde{b}_{22}\right] \ , \\
c_{12}^{(n,3)} &= 3\left[b_{n1}\tilde{b}_{11}\left(\tilde{b}_{12}\right)^2 + b_{n2}\tilde{b}_{21}\left(\tilde{b}_{22}\right)^2\right] \ , \\
c_{03}^{(n,3)} &= b_{n1}\left(\tilde{b}_{21}\right)^3 + b_{n2}\left(\tilde{b}_{22}\right)^3 \ , \qquad (12i)
\end{aligned}$$

$$\begin{aligned}
c_{20}^{(n,2)} &= -3\left[b_{n1}\left(\tilde{b}_{11}\right)^2 + b_{n2}\left(\tilde{b}_{21}\right)^2\right] \ , \\
c_{11}^{(n,2)} &= -6\left[b_{n1}\left(\tilde{b}_{11}\right)^2 + b_{n2}\left(\tilde{b}_{21}\right)^2\right] \ , \\
c_{20}^{(n,2)} &= -3\left[b_{n1}\left(\tilde{b}_{12}\right)^2 + b_{n2}\left(\tilde{b}_{22}\right)^2\right] \ , \qquad (12j)
\end{aligned}$$

$$\begin{aligned}
c_{10}^{(n,1)} &= 3\left(b_{n1}\tilde{b}_{11} + b_{n2}\tilde{b}_{11}\right) \ , \\
c_{01}^{(n,1)} &= 3\left(b_{n1}\tilde{b}_{12} + b_{n2}\tilde{b}_{12}\right) \ , \qquad (12k)
\end{aligned}$$

$$c_{00}^{(n,0)} = -a_n - 1 \ . \qquad (12l)$$

After this display of a lot of explicit formulas it is now possible to summarize the findings for this specific case, with $N = 2$ and $p = 3$: when the system of 2 *nonlinearly-coupled* recursions (2a) features 2 *cubic* polynomials $P_3(x_1, x_2)$ of the 2 dependent variables $x_1(\ell)$ and $x_2(\ell)$; each of these 2 polynomials, see (2), features $1 + 2 + 3 + 4 = 10$ coefficients:

$$\begin{aligned}
&P_3^{(n)}(x_1\,, x_2) = c_{00}^{(n,0)} + c_{10}^{(n,1)} x_1 + c_{01}^{(n,1)} x_2 \\
&+ c_{20}^{(n,2)}(x_1)^2 + c_{11}^{(n,2)} x_1 x_2 + c_{02}^{(n,2)}(x_2)^2 \\
&+ c_{30}^{(n,3)}(x_1)^3 + c_{21}^{(n,3)}(x_1)^2 x_2 + c_{12}^{(n,3)} x_1 (x_2)^2 + c_{03}^{(n,3)}(x_2)^3 \ , \qquad (13)
\end{aligned}$$

so altogether there are 20 coefficients whose values determine the $\ell$-evolution of the 2 dependent variables $x_1(\ell)$ and $x_2(\ell)$, in addition of course—in the context of the *initial-values* problem—to the 2 (*a priori arbitrary*) *initial* values $x_1(0)$ and $x_2(0)$. While the class of solutions whose behavior may be rather *explicitly* understood is restricted by the requirement that the 20 coefficients "$c$" be determined by the explicit formulas written above—see the 10 formulas (12i)-(12l)—in terms of (only!) $2+4=6$ *a priori arbitrary parameters* $a_1, a_2$ and $b_{11}, b_{12}, b_{21}, b_{22}$ (the parameters $\widetilde{b}_{11}, \widetilde{b}_{12}, \widetilde{b}_{21}, \widetilde{b}_{22}$ being of course determined by the parameters $b_{11}, b_{12}, b_{21}, b_{22}$, see **Section 2**). Thus the class of recursions (2a) (with $N=2$ and $p=3$) whose solution are now rather precisely described—as identified by the approach detailed above—is only a rather small subclass of the universe of solutions of the general system of recursions (2a) (with $N=2$ and $p=3$); and the *additional constraints* for the applicability of the findings reported above must of course also be taken into account, as implied by the additional restrictions on the *initial data* needed for the emergence of these findings (as explained above).

And now just a terse mention of the changes that would emerge when the restriction to the *very small* value $N=2$ is lifted, and the treatment given above is extended to the case of a *larger positive integer* number $N$ of variables $y_n(\ell)$ respectively $x_n(\ell)$ evolving as functions of the "ticking time" *independent* variable $\ell$. Then, for instance, for $N=3,4$ respectively 5, the numbers $\nu(N)$ of *coefficients* of each *cubic* polynomial in $N$ variables $P_3^{(n)}[x_1(\ell), ..., x_N(\ell)]$ (with $N=3,4$ respectively 5) featured by the right-hand side of *each* of the $N$ eqs. (10a-10d) would be $20, 35$ respectively 56 (which are the number of terms of a *generic* polynomial of degree 3 in $3,4$ respectively 5 variables), hence the total numbers $N\cdot\nu(N)$ of these coefficients would raise to $60, 140$ respectively 280; while the number of freely adjustable parameters $a_n$ and $b_{nm}$ playing a role in the change from the variables $y_n(\ell)$ to the variables $x_n(\ell)$ (and viceversa: see eqs.(4) and (6)) would correspondingly *grow*, albeit much more *slowly*: from $2\cdot3=6$ to $3\cdot4=12$, to $4\cdot5=20$, respectively to $5\cdot6=30$. Hence the subclass of systems of *nonlinear recursions* (2a) (with *cubic polynomial* right-hand sides) whose $\ell$-evolution is relatively easily predicted by this approach turns out to be a *smaller and smaller subclass* of the general system of such systems of recursions, as the number $N$ of *dependent* variables increases; but nevertheless a subclass rather *easily identified*, hence potentially interesting in *applicative contexts*.

Moreover a *special* case also of possible *applicative* interest is that in which the $N$ polynomials $P_p^{(n)}(x_1, ..., x_N)$ are *homogeneous* (hence they coincide with the polynomials $Q_p^{(n)}(x_1, ..., x_N)$, while all the polynomials $Q_q^{(n)}(x_1, ..., x_N)$ with $q=0,1,...,p-1$ vanish: see the eqs. (2)). Indeed in that case the numbers $\nu(N)$ of *coefficients* of each *cubic* polynomial in $N$ variables $P_3^{(n)}[x_1(\ell), ..., x_N(\ell)]$ (with $N=3,4$ respectively 5) is much reduced: to $10, 19$ respectively 33, hence for the entire system of recursions the total numbers of coefficients $N\cdot\nu(N)$ playing a role are $30, 76$ respectively 165; while, as regards the adjustable parameters $a_n$ and $b_{nm}$ (see eq. (4)), the $a_n$ would simply *disappear*, but the $b_{nm}$ would still be $N^2$, namely $9, 16$ respectively 25; so in these cases the gaps between the numbers of coefficients in the system of recursions (2) and the numbers of freely adjustable parameters in the change of variables (4) is significantly reduced.

This **Section 3** is completed by displaying *all* the relevant formulas for the special case with $N = 3$ (and of course also $p = 3$) and *homogeneous cubic* polynomials in the right-hand sides of the recursions. Then the system of recursions reads as follows:

$$x_n(\ell+1) = c_{30}^{(n,3)}[x_1(\ell)]^3 + c_{21}^{(n,3)}[x_1(\ell)]^2 x_2 + c_{12}^{(n,3)} x_1(\ell)[x_2(\ell)]^2 + c_{03}^{(n,3)}[x_2(\ell)]^3 \ ; \tag{14a}$$

and the corresponding change of variables reads as follows:

$$x_n(\ell) = b_{n1} y_1(\ell) + b_{n2} y_2(\ell) + b_{n3} y_3(\ell) \ , \tag{14b}$$

$$y_n(\ell) = \widetilde{b}_{n1} x_1(\ell) + \widetilde{b}_{n2} x_2(\ell) + \widetilde{b}_{n3} x_3(\ell) \ , \tag{14c}$$

$$\mathbf{B} = \begin{pmatrix} b_{11} & b_{12} & b_{13} \\ b_{21} & b_{22} & b_{23} \\ b_{31} & b_{32} & b_{33} \end{pmatrix} \ , \tag{14d}$$

$$\widetilde{\mathbf{B}} = \begin{pmatrix} \widetilde{b}_{11} & \widetilde{b}_{12} & \widetilde{b}_{13} \\ \widetilde{b}_{21} & \widetilde{b}_{22} & \widetilde{b}_{23} \\ \widetilde{b}_{31} & \widetilde{b}_{32} & \widetilde{b}_{33} \end{pmatrix} = \mathbf{B}^{-1} \ . \tag{14e}$$

So, a minor finding of this paper is, that it is possible to identify *rather explicitly* the evolution of the solutions of the system of *nonlinear recursions* (14a) featuring $3 \cdot 4 = 12$ *a priori arbitrary* coefficients "$c$" on their right-hand sides, provided these 12 coefficients are expressed by the 12 *explicit* formulas (12i) in terms of 9 *a priori arbitrary* parameters $b_{nm}$ (the formulas (12i) also contain the tilded quantities $\widetilde{b}_{nm}$, which are however given by well-known simple formulas in terms of the untitled parameters $b_{nm}$, as indicated above). Then, of the 12 coefficients $b_{nm}$, as many as 9 may be *arbitrarily* chosen; while the other 3 are determined by *explicit* equations (generally yielding several solutions—since these equations are *algebraic* but *not linear*), in terms of the given 9: see the eqs. (12i). The identification of the solutions $x_n(\ell)$ of the equation (14a) of course happens because the $\ell$-evolution of the 3 dependent variables $y_n(\ell)$—each of them solution of the *nonlinear* recursion (9a)— is well understood, see [5].

**4. The $\mathbf{p = 4}$ case**

In this **Section 4** a *quite analogous* treatment to that provided in the preceding **Section 3** (there, for $p = 3$) is now reported for the case with $p = 4$; but it shall be much more concise than that displayed above.

In the $p = 4$ case the system of recursions reads as follows:

$$x_n(\ell+1) = P_4^{(n)}[x_1(\ell), ..., x_4(\ell)] \ , \quad n = 1, 2, ..., N \ , \tag{15}$$

where now the $N$ polynomials $P_4^{(n)}(x_1, ..., x_4)$ are *all* of degree 4. Each of them features then 15 coefficients if $N = 2$ (but only 5 coefficients in the *homogeneous* case), hence altogether $2 \cdot 15 = 30$ *a priori arbitrary* coefficients are involved (but only $2 \cdot 5 = 10$ coefficients in the *homogeneous* case); while the corresponding *freely manageable* parameters $a_n$ and $b_{nm}$ in the *change of variables* are $2+2 \cdot 2 = 6$ (but only $2 \cdot 2 = 4$ in the *homogeneous* case). The corresponding figures for $N = 3, 4, 5$ are: for $N = 3$, $3 \cdot 35 = 105$ coefficients (but only $3 \cdot 15 = 45$ coefficients in the *homogeneous* case); for $N = 4$, $4 \cdot 70 = 280$ coefficients (but only $4 \cdot 35 = 140$ coefficients in the *homogeneous* case); for $N = 5$, $5 \cdot 126 = 630$ coefficients (but only $5 \cdot 80 = 400$ in the *homogeneous* case); while the corresponding *freely manageable* parameters $a_n$ and $b_{nm}$ in the change of variables are, for $N = 3$, 4 respectively 5, just $3 \cdot 4 = 12$, $4 \cdot 5 = 20$ respectively $5 \cdot 6 = 30$ (but only 9, 16 respectively 25 in the *homogeneous* case). These figures provide some glimpse of the potential usefulness of these findings in eventual *applicative* contexts. The relevant formulas are *analogous*—although of course progressively somewhat *more complicated*—than those *explicitly* displayed above; but the route to obtain them is plain.

## 5. Final remarks

Extensions of the findings reported above to *larger* values of $p$ and $N$ are not reported in this paper.

*Other* extensions of the simple approach used in this paper may of course be envisaged, to identify systems of *nonlinear first-order recursions*—or even of *more general recursions*—featuring solutions which evolve in the ticking-time variable $\ell$ *rather predictably* (for instance being *periodic* or *asymptotically periodic*); possibly by replacing the *linear* change of variable (4) with some other *nonlinear* yet still *easily manageable* change of variables. Maybe I will myself make some progress in this directions in the future; but it would be also desirable that other researchers try and do so; since it seems to me that the *applicative* developments of these findings are likely to be relevant. In such cases, I would be much grateful to be kept promptly informed (say, by *e-mail*) of any such significant development.